\documentclass[aps,prb,twocolumn,floatfix,showpacs,letterpaper]{revtex4}

\usepackage{amsmath}
\usepackage{graphicx}

\begin{document}

\title{Strong coupling of localized plasmons and molecular excitons in nanostructured silver films}

\author{N. I. Cade}
\email{nicholas.cade@kcl.ac.uk}
\author{T. Ritman-Meer}
\author{D. Richards}
\affiliation{Department of Physics, King's College London, Strand, London WC2R 2LS, UK}


\begin{abstract}
We report on the resonant coupling between localized surface plasmon resonances (LSPRs) in nanostructured Ag films, and an adsorbed monolayer of Rhodamine 6G dye. Hybridization of the plasmons and molecular excitons creates new coupled polaritonic modes, which have been tuned by varying the LSPR wavelength. The resulting polariton dispersion curve shows an anticrossing behavior which is very well fit by a simple  coupled-oscillator Hamiltonian, giving a giant Rabi-splitting energy of $\sim$400~meV. The strength of this coupling is shown to be proportional to the square root of the molecular density. The Raman spectra of R6G on these films show an enhancement of many orders of magnitude due to surface enhanced scattering mechanisms; we find a maximum signal when a polariton mode lies in the middle of the Stokes shifted emission band.
\end{abstract}
\maketitle

There is currently considerable interest in the interaction between excitonic and photonic states, as a means of modifying the photophysical properties of a system. Potential novel applications include lasers,\cite{Christopoulos2007} optical switches,\cite{Bajoni2008} and sensors.\cite{Lee2007} In microcavities, mixing of exciton and photon modes leads to the formation of new polaritonic states, and has been observed in both organic\cite{Lidzey1998,Hobson2002,Takada2003} and inorganic systems.\cite{Khitrova2006} More recently, coupling has been observed between excitonic and plasmonic states for semiconductor heterostructures.\cite{Bellessa2008,Vasa2008} Localized plasmons are the subject of many current investigations, as they can dramatically alter the optical properties of a locally situated molecule: enhancement and confinement of the excitation field has important consequences in surface enhanced Raman scattering (SERS).\cite{nie97} Furthermore, localized surface plasmon resonances (LSPRs) can be engineered to produce large modifications in fluorescence intensity and lifetime.\cite{Ritman-Meer2007,Cade09a} Due to this, the interaction between localized plasmon modes and excitonic states has been studied recently for a variety of nanostructured systems: these include  nanoparticles,\cite{Zhao2007,Fofang2008} nanorods,\cite{Ni2008} nanovoids,\cite{Sugawara2006} and subwavelength hole arrays.\cite{Dintinger2005}  For all these systems, strong coupling is manifested as an anticrossing behavior in the dispersion curve of the plasmon mode at the energy of the uncoupled exciton mode, indicating the formation of a hybridized exciton-plasmon polariton state; the resulting mode splitting is determined by the coupling strength of the two systems.

In this work, we report on the resonant coupling between LSPRs in nanostructured silver films (NSFs), and two different excitonic states in an adsorbed dye, and we demonstrate the importance of this mechanism for SERS. The coupling strength was tuned by varying the LSPR wavelength from 450 to 750 nm; the resulting exciton-plasmon polariton peak positions are very well fit by  a three-coupled-oscillator Hamiltonian, which gives a Rabi-splitting energy comparable to the largest values reported to date.  Raman spectra have been taken for each film at two different wavelengths, and in both cases we find a maximum signal enhancement when the middle of the Stokes shifted emission band is resonant with a polariton mode rather than the plasmonic or molecular absorption peak.

\begin{figure}[tb]
 \begin{center} \includegraphics[width=7.6cm]{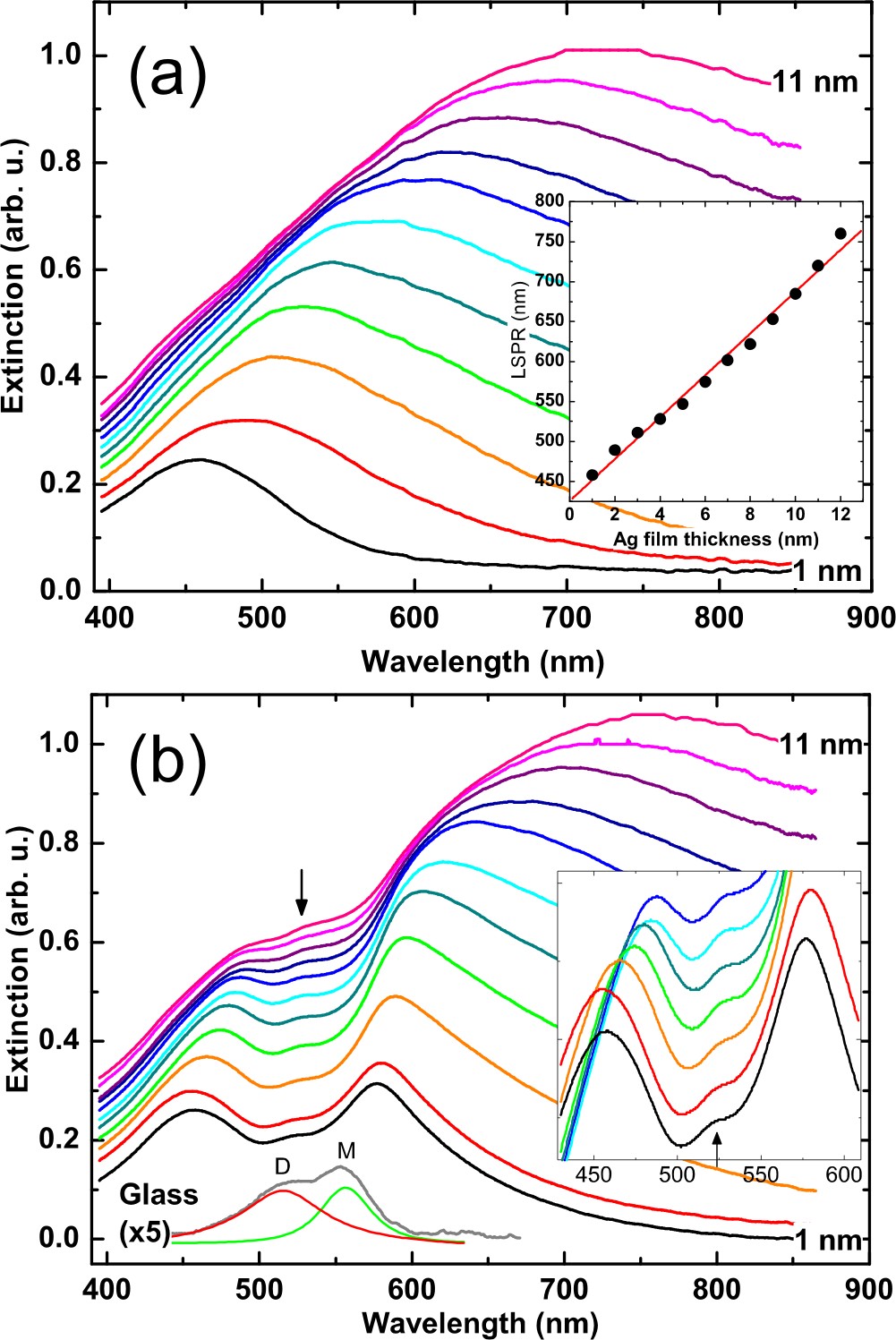} \end{center}
  \caption{(Color online)(a) Extinction spectra from bare NSFs of increasing thickness. Inset: LSPR position vs nominal film thicknesses. The solid line is a linear fit. (b) Corresponding extinction spectra after deposition of R6G on the films. The short wavelength region is enlarged in the inset showing the middle polariton mode (arrow). The bottom spectrum is R6G on glass with Gaussian fits indicating monomer (M) and dimer (D) formation.} \label{extinction}
\end{figure}

Glass coverslips were cleaned in an acid piranha solution (3:1  H$_2$SO$_4$ / H$_2$O$_2$) for 60 minutes, and then rinsed thoroughly in deionized water. Thin silver films were prepared by thermal deposition of Ag (99.99$\%$ purity) in a vacuum chamber at 10$^{-6}$ Torr: on one coverslip the nominal thickness of silver deposited was varied from 1--12 nm, in 1 nm increments, as measured with a quartz crystal oscillator. The films were characterized using atomic force microscopy (Veeco Explorer): all of the films have an irregular nanostructured surface, with the average particle size increasing with nominal thickness. Extinction spectra (Perkin Elmer UV900) were taken for each NSF, and are shown unscaled in  Fig.\ \ref{extinction}(a). For the films studied here, the position of the LSPR will depend on the size, shape, and density of the Ag particles;\cite{Gupta2002} we find that the LSPR position increases approximately linearly with nominal film thickness, as shown in the inset of Fig.\ \ref{extinction}(a). Rhodamine 6G (R6G) dye was deposited onto all the films simultaneously by vacuum sublimation to produce a uniform coverage of approximately monolayer thickness. Extinction spectra were acquired again for each film and for the R6G on bare glass; these are shown in  Fig.\ \ref{extinction}(b). The R6G on glass spectrum can be deconvolved into two distinct peaks at 555 and 512 nm, which are assigned to monomer and H-dimer excitons, respectively.\cite{Selwyn1972,Zhao2007} For the molecular densities investigated here, there is no evidence of J-dimer formation. Comparing  Figs.\ \ref{extinction}(a) and (b), we see a clear modification in the absorption of the combined system which cannot be accounted for with a simple combination of the constituent absorbers.\cite{Glass1980} This is due to strong coupling of the molecular and plasmonic states, resulting in the formation of hybridized exciton-plasmon polariton modes. The inset in  Fig.\ \ref{extinction}(b) shows clearly there are three polariton peaks which red-shift with increasing film thickness.

\begin{figure}[tb]
 \begin{center} \includegraphics[width=8cm]{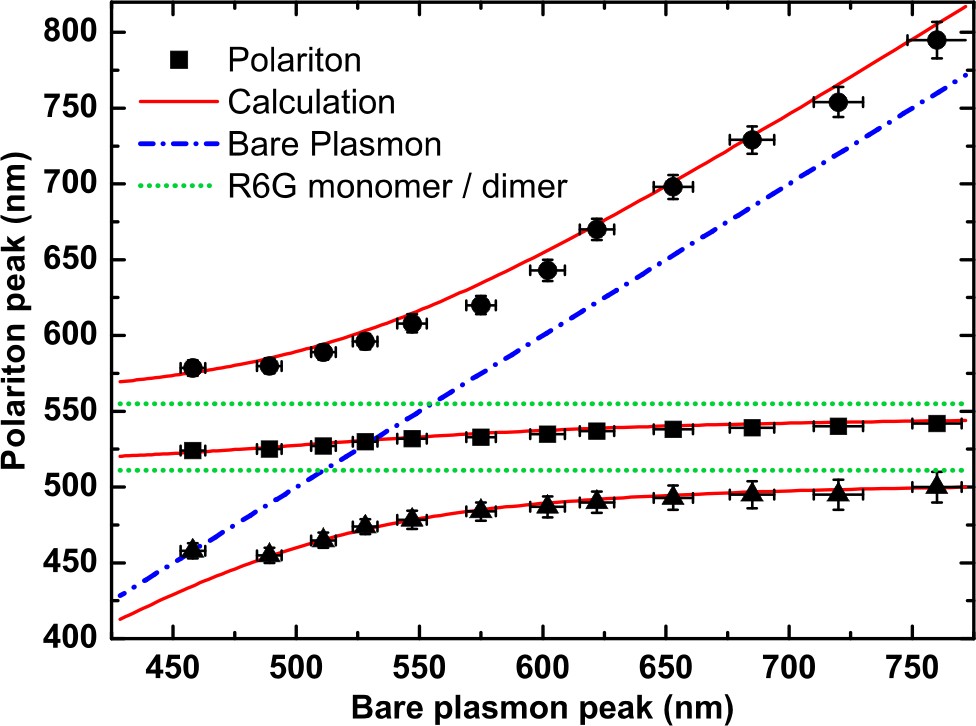} \end{center}
  \caption{(Color online) Position of the peaks in Fig.\ \ref{extinction}(b) plotted against the corresponding bare LSPR position (black symbols). Also shown are the R6G exciton peaks (dotted green), and the polariton dispersion curves calculated from Eq.\ 1 (solid red).} \label{calc}
\end{figure}

Figure \ref{calc} shows the position of the peaks in Fig.\ \ref{extinction}(b) plotted against the bare LSPR position, for each corresponding film thickness; the dashed and dotted lines are the plasmon and exciton positions, respectively. The data show a clear anticrossing behavior at the energies of both excitons. To better understand the coupling mechanism, the polariton dispersions were calculated using a simplified coupled oscillator model:\cite{Bellessa2008,Wurtz2007} we consider a physical system with Hamiltonian $H_{\textrm{0}}$ and eigenstates $\mid p \rangle$, $\mid m \rangle$, and $\mid d \rangle$  associated with the plasmon resonance and the R6G  monomer and dimer excitons, respectively. The corresponding eigenvalues of the system are $E_{\textrm{p}}$, $E_{\textrm{m}}$, and $E_{\textrm{d}}$, respectively. We assume that both excitonic eigenstates have an equal time-independent coupling $\Delta$ to the plasmonic eigenstate, and we ignore any homogeneous broadening effects; hence, the Hamiltonian can be rewritten as $H =  H_{\textrm{0}} + \Delta $. This gives
\begin{equation}
H=    \begin{pmatrix}
      E_{\textrm{p}}^{'} & \Delta & \Delta \\
      \Delta & E_{\textrm{m}}^{'} & 0 \\
      \Delta & 0 & E_{\textrm{d}}^{'} \\
    \end{pmatrix}
  \end{equation}
where $E_{\textrm{p}}^{'}$, $E_{\textrm{m}}^{'}$, and $E_{\textrm{d}}^{'}$ are the uncoupled eigenvalues modified due to a change in the dielectric environment. For the molecular states, this change in energy only depends on the geometry of the interface rather than the material properties of the substrate itself;\cite{Wurtz2007} hence, we have taken $E_{\textrm{m}}^{'}$ and $E_{\textrm{d}}^{'}$ directly from the peak fits of R6G on glass in  Fig.\ \ref{extinction}(b). The eigenvalues of Eq.\ 1 were obtained numerically by assuming $E_{\textrm{p}}^{'} \simeq E_{\textrm{p}}$, and have been fitted to the data in  Fig.\ \ref{calc} by varying only $\Delta$. We find excellent agreement with the experimental results over the whole range of wavelengths, with a coupling strength of 190 meV corresponding to a Rabi splitting of 2$\Delta = 380$ meV. This giant splitting is comparable to those reported for organic microcavities;\cite{Hobson2002,Takada2003} it originates from the high oscillator strength of the dye and also the large local field amplitudes generated by surface plasmons.\cite{Dintinger2005} In our case the plasmons are localized in individual Ag nanoparticles of the NSF, which will generate a much higher local field intensity than at the surface of a smooth film.\cite{Markel1999}

\begin{figure}[tb]
 \begin{center} \includegraphics[width=8.15cm]{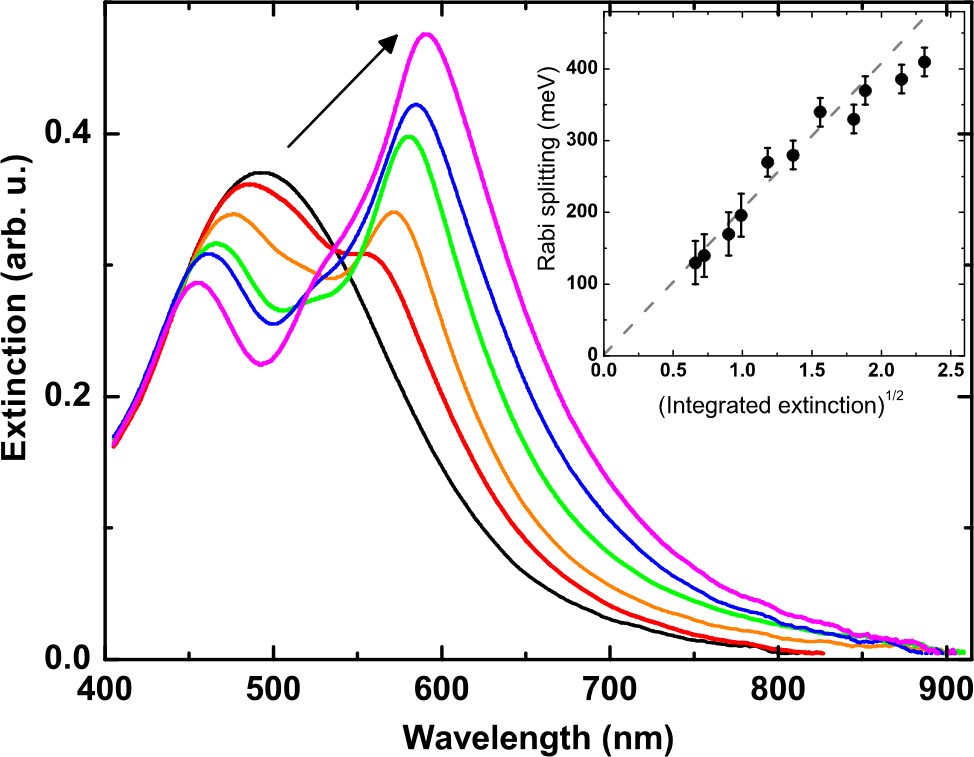} \end{center}
  \caption{(Color online) Extinction spectra from a 3~nm Ag film with increasing coverage of R6G, as indicated by the arrow. Inset: corresponding Rabi splitting 2$\Delta$, calculated from Eq.\ 1, as a function of the square root of the integrated R6G extinction on glass (arbitrary units). The dashed line is a guide to the eye.} \label{conc}
\end{figure}

The Rabi splitting 2$\Delta$ is expected to vary approximately as the square root of the molecular absorption.\cite{Norris,Lidzey1998,Dintinger2005} To verify this, a uniform nominal 3~nm NSF was covered with increasing amounts of R6G; the resulting extinction spectra are shown in Fig.\ \ref{conc}. For each level of coverage, the Rabi splitting was calculated from the polariton peak positions using the solutions of Eq.\ 1, and the integrated extinction was measured for the R6G on bare glass. The inset in Fig.\ \ref{conc} shows there is a very good agreement with the predict square root dependence. For a system of isolated nanoparticles, the spatial extent of the LSPR field is of the order of the dimensions of the nanoparticles; thus, for low coverages, increasing the density of absorbers increases the number of molecules which can couple to each LSPR.  The deviation observed at the highest molecular densities is probably due to saturation of electromagnetic `hotspots' on the NSF above a surface coverage of approximately a monolayer.

The principal requirement to observe strong coupling is that the linewidths of both the excitons and plasmon must be less that the Rabi-splitting energy.\cite{Norris} For the NSFs investigated here, the measured LSPR linewidth will be mainly due to the inhomogeneous size and shape distribution of constituent nanoparticles. However, we assume that interactions with the covering molecules will be dominated by those nanoparticle in the peak of the distribution, with each individual nanoparticle having an intrinsic linewidth much smaller than that measured.\cite{Noguez2007} For the R6G, the peak fits in Fig.\ \ref{extinction}(b) give linewidths of $\lesssim$160 meV, which is consistent with the minimum value of the Rabi splitting for which strong coupling is observed in Fig.\ \ref{conc}.


\begin{figure}[tb]
\begin{center}
\includegraphics[width=7.8cm]{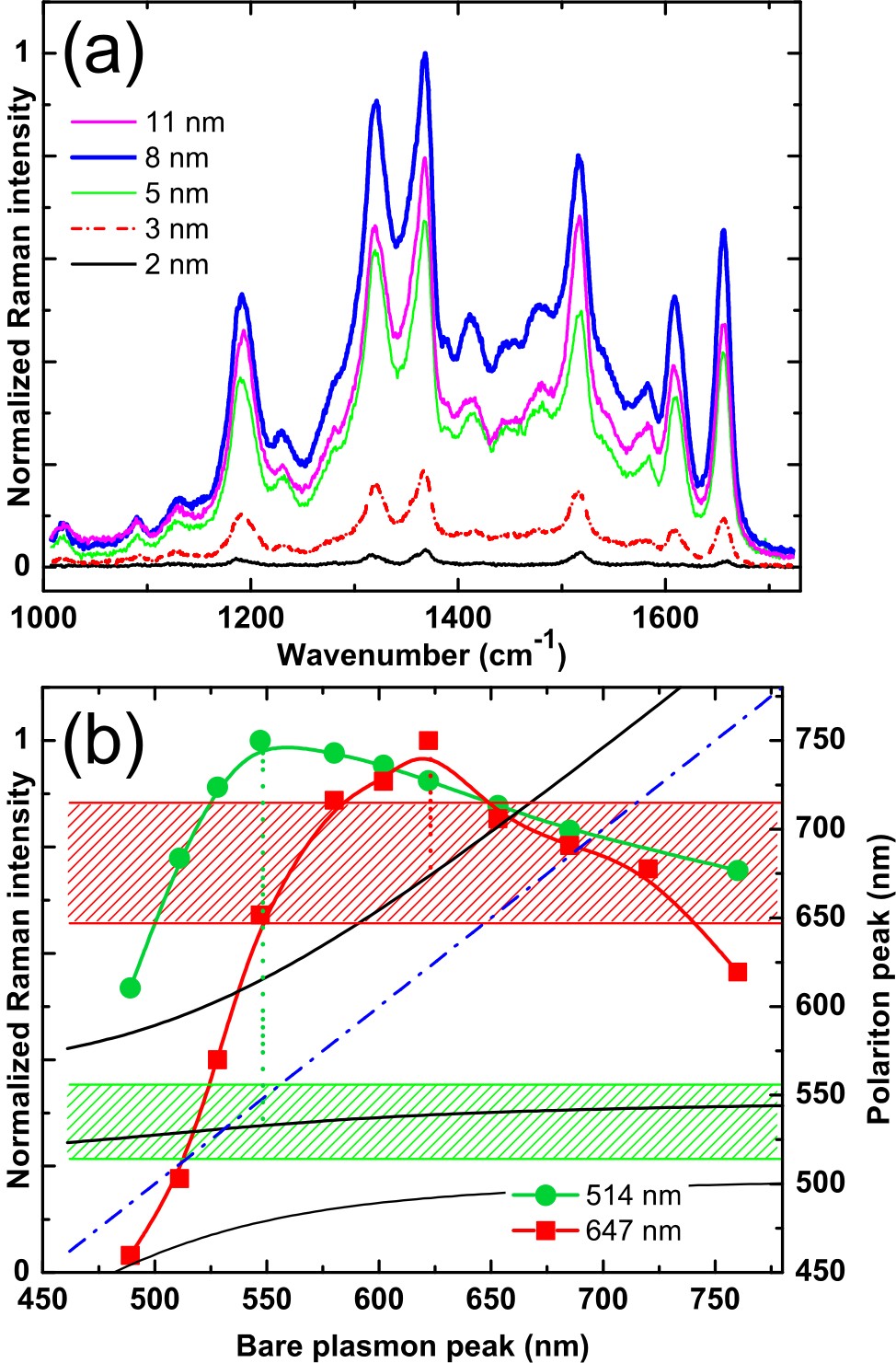}
\end{center}
\caption{(Color online)(a)  Raman spectra from R6G on NSFs of increasing thickness (647 nm excitation). (b) Left axis: integrated Raman intensity from the NSFs as a function of the bare plasmon peak position. Excitation was at 514 nm (green circles) and 647 nm (red squares). Lines are guides to the eye. Right axis: polariton peak positions from  Fig.\ \ref{calc} (solid black lines), and bare LSPR position (dashed blue line). The shaded areas correspond to the wavelength range between the incident laser and Raman peaks at 1350 cm$^{-1}$. Maximum SERS enhancement occurs when a polariton mode is in the middle of this range (dotted lines).}\label{raman}
\end{figure}

To investigate further the effects of mode hybridization in the NSFs, spatially averaged Raman spectra were acquired (Renishaw 1000 microscope) for R6G on the varying thickness films described above. Excitation was at 514 nm (22 $\mu$W) and  647 nm (138 $\mu$W) with a 20$\times$ objective lens (Leica, 0.4 NA). Background-subtracted spectra from selected films are shown in Fig.\ \ref{raman}(a) for 647 nm excitation. No Raman signal was detectable from the R6G deposited directly onto glass. All films show a large enhancement in the R6G Raman signal due to SERS,\cite{campion98,Moskovits2005} with a strong dependence on nominal thickness and hence LSPR position. For each film, Fig.\ \ref{raman}(b) shows the integrated intensity of the Raman peaks at $\sim$1350 cm$^{-1}$ as a function of the corresponding bare LSPR position, for both excitation wavelengths. Using the total integrated intensity between 1000--1700 cm$^{-1}$ gives a similar trend. For 514 nm excitation, a maximum enhancement occurs for the 5 nm film with an LSPR at 550 nm, whereas for 647 nm excitation, the largest signal is from the 8 nm film with an LSPR at 620 nm.

In a weakly coupled system, the dominant SERS enhancement mechanism is usually amplification of the incident and scattered electric fields due to e.g.\ localized plasmons; the greatest enhancement has been found to occur when there is a maximum overlap between the LSPR and both laser and Raman lines.\cite{Felidj2003,McFarland2005} The dashed line in Fig.\ \ref{raman}(b) shows that this would occur for the 4 nm and 10 nm films with 514 nm and 647 nm excitation, respectively. When compared with the polariton peak positions in  Fig.\ \ref{calc},  Fig.\ \ref{raman}(b) shows that in a strongly coupled system the largest SERS signal is produced when the polariton mode is halfway between the incident laser and the Raman emission line. There is a 30-fold change in SERS signal for 647 nm excitation, as the upper polariton mode crosses the Raman enhancement region rapidly with LSPR tuning. In contrast, for 514 nm excitation there is only a two-fold change in SERS signal over the whole wavelength range, as the central polariton mode varies little with tuning. A similar analysis of a 450 cm$^{-1}$ Raman peak close to the 647 nm excitation line gives a maximum signal for the 7 nm film (LSPR $\sim$600 nm); this is consistent with the above interpretation as the upper polariton mode is closest to the laser line for this film.

In addition to the electromagnetic enhancement mechanism, SERS effects are also caused by a chemical enhancement.\cite{Moskovits2005,Persson2006} This mechanism has been attributed to charge transfer between the metal and molecular orbitals, and arises from a mixing of the molecular and metallic states; this can result in an additional $>$10$^2$ increase in scattering signal. In a strongly coupled system with a mesoscopic plasmonic nanostructure, such as we observe here, the two mechanisms are essentially indistinguishable:\cite{Pettinger1986} this is manifested as a resonant Raman enhancement at the wavelength of the coupled exciton-plasmon polariton mode rather than at the molecular or LSPR absorption maximum. The nature of the different enhancement mechanisms in SERS is still the subject of many investigations; thus, these experimental results are of particular importance in the development of a general microscopic theory for SERS.

In conclusion, nanostructured silver films have been fabricated in which the plasmon resonance can be easily tuned by varying the nominal film thickness. We have shown that strong coupling occurs between localized surface plasmons and molecular excitonic states in an adsorbed dye. The dispersion of the resulting exciton-plasmon polaritonic states is in excellent agreement with a coupled-oscillator model which gives a giant Rabi splitting of $\sim$400~meV. We have verified that the coupling strength depends on the square root of the density of the absorber.  SERS spectra indicate that in a strongly coupled system the greatest signal enhancement occurs when a hybridized polariton mode lies midway between the laser excitation and the Raman emission lines.

This work was supported by the EPSRC (UK).


\begin{thebibliography}{30}
\expandafter\ifx\csname natexlab\endcsname\relax\def\natexlab#1{#1}\fi
\expandafter\ifx\csname bibnamefont\endcsname\relax
  \def\bibnamefont#1{#1}\fi
\expandafter\ifx\csname bibfnamefont\endcsname\relax
  \def\bibfnamefont#1{#1}\fi
\expandafter\ifx\csname citenamefont\endcsname\relax
  \def\citenamefont#1{#1}\fi
\expandafter\ifx\csname url\endcsname\relax
  \def\url#1{\texttt{#1}}\fi
\expandafter\ifx\csname urlprefix\endcsname\relax\def\urlprefix{URL }\fi
\providecommand{\bibinfo}[2]{#2}
\providecommand{\eprint}[2][]{\url{#2}}

\bibitem[{\citenamefont{Christopoulos et~al.}(2007)\citenamefont{Christopoulos,
  von Hogersthal, Grundy, Lagoudakis, Kavokin, Baumberg, Christmann, Butte,
  Feltin, Carlin et~al.}}]{Christopoulos2007}
\bibinfo{author}{\bibfnamefont{S.}~\bibnamefont{Christopoulos}},
  \bibinfo{author}{\bibfnamefont{G.~B.~H.} \bibnamefont{von Hogersthal}},
  \bibinfo{author}{\bibfnamefont{A.~J.~D.} \bibnamefont{Grundy}},
  \bibinfo{author}{\bibfnamefont{P.~G.} \bibnamefont{Lagoudakis}},
  \bibinfo{author}{\bibfnamefont{A.~V.} \bibnamefont{Kavokin}},
  \bibinfo{author}{\bibfnamefont{J.~J.} \bibnamefont{Baumberg}},
  \bibinfo{author}{\bibfnamefont{G.}~\bibnamefont{Christmann}},
  \bibinfo{author}{\bibfnamefont{R.}~\bibnamefont{Butte}},
  \bibinfo{author}{\bibfnamefont{E.}~\bibnamefont{Feltin}},
  \bibinfo{author}{\bibfnamefont{J.-F.} \bibnamefont{Carlin}},
  \bibnamefont{and} \bibinfo{author}{\bibfnamefont{N.}~\bibnamefont{Grandjean}},
  \bibinfo{journal}{Phys. Rev. Lett.}
  \textbf{\bibinfo{volume}{98}}, \bibinfo{pages}{126405}
  (\bibinfo{year}{2007}).

\bibitem[{\citenamefont{Bajoni et~al.}(2008)\citenamefont{Bajoni, Semenova,
  Lemaitre, Bouchoule, Wertz, Senellart, Barbay, Kuszelewicz, and
  Bloch}}]{Bajoni2008}
\bibinfo{author}{\bibfnamefont{D.}~\bibnamefont{Bajoni}},
  \bibinfo{author}{\bibfnamefont{E.}~\bibnamefont{Semenova}},
  \bibinfo{author}{\bibfnamefont{A.}~\bibnamefont{Lemaitre}},
  \bibinfo{author}{\bibfnamefont{S.}~\bibnamefont{Bouchoule}},
  \bibinfo{author}{\bibfnamefont{E.}~\bibnamefont{Wertz}},
  \bibinfo{author}{\bibfnamefont{P.}~\bibnamefont{Senellart}},
  \bibinfo{author}{\bibfnamefont{S.}~\bibnamefont{Barbay}},
  \bibinfo{author}{\bibfnamefont{R.}~\bibnamefont{Kuszelewicz}},
  \bibnamefont{and} \bibinfo{author}{\bibfnamefont{J.}~\bibnamefont{Bloch}},
  \bibinfo{journal}{Phys. Rev. Lett.} \textbf{\bibinfo{volume}{101}},
  \bibinfo{pages}{266402} (\bibinfo{year}{2008}).

\bibitem[{\citenamefont{Lee et~al.}(2007)\citenamefont{Lee, Hernandez, Lee,
  Govorov, and Kotov}}]{Lee2007}
\bibinfo{author}{\bibfnamefont{J.}~\bibnamefont{Lee}},
  \bibinfo{author}{\bibfnamefont{P.}~\bibnamefont{Hernandez}},
  \bibinfo{author}{\bibfnamefont{J.}~\bibnamefont{Lee}},
  \bibinfo{author}{\bibfnamefont{A.~O.} \bibnamefont{Govorov}},
  \bibnamefont{and} \bibinfo{author}{\bibfnamefont{N.~A.} \bibnamefont{Kotov}},
  \bibinfo{journal}{Nat Mater} \textbf{\bibinfo{volume}{6}},
  \bibinfo{pages}{291} (\bibinfo{year}{2007}).

\bibitem[{\citenamefont{Lidzey et~al.}(1998)\citenamefont{Lidzey, Bradley,
  Skolnick, Virgili, Walker, and Whittaker}}]{Lidzey1998}
\bibinfo{author}{\bibfnamefont{D.~G.} \bibnamefont{Lidzey}},
  \bibinfo{author}{\bibfnamefont{D.~D.~C.} \bibnamefont{Bradley}},
  \bibinfo{author}{\bibfnamefont{M.~S.} \bibnamefont{Skolnick}},
  \bibinfo{author}{\bibfnamefont{T.}~\bibnamefont{Virgili}},
  \bibinfo{author}{\bibfnamefont{S.}~\bibnamefont{Walker}}, \bibnamefont{and}
  \bibinfo{author}{\bibfnamefont{D.~M.} \bibnamefont{Whittaker}},
  \bibinfo{journal}{Nature} \textbf{\bibinfo{volume}{395}}, \bibinfo{pages}{53}
  (\bibinfo{year}{1998}).

\bibitem[{\citenamefont{Hobson et~al.}(2002)\citenamefont{Hobson, Barnes,
  Lidzey, Gehring, Whittaker, Skolnick, and Walker}}]{Hobson2002}
\bibinfo{author}{\bibfnamefont{P.~A.} \bibnamefont{Hobson}},
  \bibinfo{author}{\bibfnamefont{W.~L.} \bibnamefont{Barnes}},
  \bibinfo{author}{\bibfnamefont{D.~G.} \bibnamefont{Lidzey}},
  \bibinfo{author}{\bibfnamefont{G.~A.} \bibnamefont{Gehring}},
  \bibinfo{author}{\bibfnamefont{D.~M.} \bibnamefont{Whittaker}},
  \bibinfo{author}{\bibfnamefont{M.~S.} \bibnamefont{Skolnick}},
  \bibnamefont{and} \bibinfo{author}{\bibfnamefont{S.}~\bibnamefont{Walker}},
  \bibinfo{journal}{Appl. Phys. Lett.} \textbf{\bibinfo{volume}{81}},
  \bibinfo{pages}{3519} (\bibinfo{year}{2002}).

\bibitem[{\citenamefont{Takada et~al.}(2003)\citenamefont{Takada, Kamata, and
  Bradley}}]{Takada2003}
\bibinfo{author}{\bibfnamefont{N.}~\bibnamefont{Takada}},
  \bibinfo{author}{\bibfnamefont{T.}~\bibnamefont{Kamata}}, \bibnamefont{and}
  \bibinfo{author}{\bibfnamefont{D.~D.~C.} \bibnamefont{Bradley}},
  \bibinfo{journal}{Appl. Phys. Lett.} \textbf{\bibinfo{volume}{82}},
  \bibinfo{pages}{1812} (\bibinfo{year}{2003}).

\bibitem[{\citenamefont{Khitrova et~al.}(2006)\citenamefont{Khitrova, Gibbs,
  Kira, Koch, and Scherer}}]{Khitrova2006}
\bibinfo{author}{\bibfnamefont{G.}~\bibnamefont{Khitrova}},
  \bibinfo{author}{\bibfnamefont{H.~M.} \bibnamefont{Gibbs}},
  \bibinfo{author}{\bibfnamefont{M.}~\bibnamefont{Kira}},
  \bibinfo{author}{\bibfnamefont{S.~W.} \bibnamefont{Koch}}, \bibnamefont{and}
  \bibinfo{author}{\bibfnamefont{A.}~\bibnamefont{Scherer}},
  \bibinfo{journal}{Nat Phys} \textbf{\bibinfo{volume}{2}}, \bibinfo{pages}{81}
  (\bibinfo{year}{2006}).

\bibitem[{\citenamefont{Bellessa et~al.}(2008)\citenamefont{Bellessa, Symonds,
  Meynaud, Plenet, Cambril, Miard, Ferlazzo, and Lemaitre}}]{Bellessa2008}
\bibinfo{author}{\bibfnamefont{J.}~\bibnamefont{Bellessa}},
  \bibinfo{author}{\bibfnamefont{C.}~\bibnamefont{Symonds}},
  \bibinfo{author}{\bibfnamefont{C.}~\bibnamefont{Meynaud}},
  \bibinfo{author}{\bibfnamefont{J.~C.} \bibnamefont{Plenet}},
  \bibinfo{author}{\bibfnamefont{E.}~\bibnamefont{Cambril}},
  \bibinfo{author}{\bibfnamefont{A.}~\bibnamefont{Miard}},
  \bibinfo{author}{\bibfnamefont{L.}~\bibnamefont{Ferlazzo}}, \bibnamefont{and}
  \bibinfo{author}{\bibfnamefont{A.}~\bibnamefont{Lemaitre}},
  \bibinfo{journal}{Phys. Rev. B} \textbf{\bibinfo{volume}{78}},
  \bibinfo{pages}{205326} (\bibinfo{year}{2008}).

\bibitem[{\citenamefont{Vasa et~al.}(2008)\citenamefont{Vasa, Pomraenke,
  Schwieger, Mazur, Kunets, Srinivasan, Johnson, Kihm, Kim, Runge
  et~al.}}]{Vasa2008}
\bibinfo{author}{\bibfnamefont{P.}~\bibnamefont{Vasa}},
  \bibinfo{author}{\bibfnamefont{R.}~\bibnamefont{Pomraenke}},
  \bibinfo{author}{\bibfnamefont{S.}~\bibnamefont{Schwieger}},
  \bibinfo{author}{\bibfnamefont{Y.~I.} \bibnamefont{Mazur}},
  \bibinfo{author}{\bibfnamefont{V.}~\bibnamefont{Kunets}},
  \bibinfo{author}{\bibfnamefont{P.}~\bibnamefont{Srinivasan}},
  \bibinfo{author}{\bibfnamefont{E.}~\bibnamefont{Johnson}},
  \bibinfo{author}{\bibfnamefont{J.~E.} \bibnamefont{Kihm}},
  \bibinfo{author}{\bibfnamefont{D.~S.} \bibnamefont{Kim}},
  \bibinfo{author}{\bibfnamefont{E.}~\bibnamefont{Runge}},
  \bibinfo{author}{\bibfnamefont{G.}~\bibnamefont{Salamo}},
  \bibnamefont{and} \bibinfo{author}{\bibfnamefont{C.}~\bibnamefont{Lienau}},
  \bibinfo{journal}{Phys. Rev. Lett.}
  \textbf{\bibinfo{volume}{101}}, \bibinfo{pages}{116801}
  (\bibinfo{year}{2008}).

\bibitem[{\citenamefont{Nie and Emory}(1997)}]{nie97}
\bibinfo{author}{\bibfnamefont{S.}~\bibnamefont{Nie}} \bibnamefont{and}
  \bibinfo{author}{\bibfnamefont{S.~R.} \bibnamefont{Emory}},
  \bibinfo{journal}{Science} \textbf{\bibinfo{volume}{275}},
  \bibinfo{pages}{1102} (\bibinfo{year}{1997}).

\bibitem[{\citenamefont{Ritman-Meer et~al.}(2007)\citenamefont{Ritman-Meer,
  Cade, and Richards}}]{Ritman-Meer2007}
\bibinfo{author}{\bibfnamefont{T.}~\bibnamefont{Ritman-Meer}},
  \bibinfo{author}{\bibfnamefont{N.~I.} \bibnamefont{Cade}}, \bibnamefont{and}
  \bibinfo{author}{\bibfnamefont{D.}~\bibnamefont{Richards}},
  \bibinfo{journal}{Appl. Phys. Lett.} \textbf{\bibinfo{volume}{91}},
  \bibinfo{pages}{123122} (\bibinfo{year}{2007}).

\bibitem[{\citenamefont{Cade et~al.}(2009)\citenamefont{Cade, Culfaz, Eligal,
  Ritman-Meer, Huang, Festy, and Richards}}]{Cade09a}
\bibinfo{author}{\bibfnamefont{N.~I.} \bibnamefont{Cade}},
  \bibinfo{author}{\bibfnamefont{F.}~\bibnamefont{Culfaz}},
  \bibinfo{author}{\bibfnamefont{L.}~\bibnamefont{Eligal}},
  \bibinfo{author}{\bibfnamefont{T.}~\bibnamefont{Ritman-Meer}},
  \bibinfo{author}{\bibfnamefont{F.-M.} \bibnamefont{Huang}},
  \bibinfo{author}{\bibfnamefont{F.}~\bibnamefont{Festy}}, \bibnamefont{and}
  \bibinfo{author}{\bibfnamefont{D.}~\bibnamefont{Richards}},
  \bibinfo{journal}{Nanobiotechnol.},
  \bibinfo{doi}{10.1007/s12030-009-9020-x}
  (\bibinfo{year}{to be published 2009}).

\bibitem[{\citenamefont{Zhao et~al.}(2007)\citenamefont{Zhao, Jensen, Sung,
  Zou, Schatz, and VanDuyne}}]{Zhao2007}
\bibinfo{author}{\bibfnamefont{J.}~\bibnamefont{Zhao}},
  \bibinfo{author}{\bibfnamefont{L.}~\bibnamefont{Jensen}},
  \bibinfo{author}{\bibfnamefont{J.}~\bibnamefont{Sung}},
  \bibinfo{author}{\bibfnamefont{S.}~\bibnamefont{Zou}},
  \bibinfo{author}{\bibfnamefont{G.}~\bibnamefont{Schatz}}, \bibnamefont{and}
  \bibinfo{author}{\bibfnamefont{R.}~\bibnamefont{VanDuyne}},
  \bibinfo{journal}{J. Am. Chem. Soc.} \textbf{\bibinfo{volume}{129}},
  \bibinfo{pages}{7647} (\bibinfo{year}{2007}).

\bibitem[{\citenamefont{Fofang et~al.}(2008)\citenamefont{Fofang, Park,
  Neumann, Mirin, Nordlander, and Halas}}]{Fofang2008}
\bibinfo{author}{\bibfnamefont{N.~T.} \bibnamefont{Fofang}},
  \bibinfo{author}{\bibfnamefont{T.-H.} \bibnamefont{Park}},
  \bibinfo{author}{\bibfnamefont{O.}~\bibnamefont{Neumann}},
  \bibinfo{author}{\bibfnamefont{N.~A.} \bibnamefont{Mirin}},
  \bibinfo{author}{\bibfnamefont{P.}~\bibnamefont{Nordlander}},
  \bibnamefont{and} \bibinfo{author}{\bibfnamefont{N.~J.} \bibnamefont{Halas}},
  \bibinfo{journal}{Nano Lett.} \textbf{\bibinfo{volume}{8}},
  \bibinfo{pages}{3481} (\bibinfo{year}{2008}).

\bibitem[{\citenamefont{Ni et~al.}(2008)\citenamefont{Ni, Yang, Chen, Li, and
  Wang}}]{Ni2008}
\bibinfo{author}{\bibfnamefont{W.}~\bibnamefont{Ni}},
  \bibinfo{author}{\bibfnamefont{Z.}~\bibnamefont{Yang}},
  \bibinfo{author}{\bibfnamefont{H.}~\bibnamefont{Chen}},
  \bibinfo{author}{\bibfnamefont{L.}~\bibnamefont{Li}}, \bibnamefont{and}
  \bibinfo{author}{\bibfnamefont{J.}~\bibnamefont{Wang}}, \bibinfo{journal}{J.
  Am. Chem. Soc.} \textbf{\bibinfo{volume}{130}}, \bibinfo{pages}{6692}
  (\bibinfo{year}{2008}).

\bibitem[{\citenamefont{Sugawara et~al.}(2006)\citenamefont{Sugawara, Kelf,
  Baumberg, Abdelsalam, and Bartlett}}]{Sugawara2006}
\bibinfo{author}{\bibfnamefont{Y.}~\bibnamefont{Sugawara}},
  \bibinfo{author}{\bibfnamefont{T.~A.} \bibnamefont{Kelf}},
  \bibinfo{author}{\bibfnamefont{J.~J.} \bibnamefont{Baumberg}},
  \bibinfo{author}{\bibfnamefont{M.~E.} \bibnamefont{Abdelsalam}},
  \bibnamefont{and} \bibinfo{author}{\bibfnamefont{P.~N.}
  \bibnamefont{Bartlett}}, \bibinfo{journal}{Phys. Rev. Lett.}
  \textbf{\bibinfo{volume}{97}}, \bibinfo{pages}{266808}
  (\bibinfo{year}{2006}).

\bibitem[{\citenamefont{Dintinger et~al.}(2005)\citenamefont{Dintinger, Klein,
  Bustos, Barnes, and Ebbesen}}]{Dintinger2005}
\bibinfo{author}{\bibfnamefont{J.}~\bibnamefont{Dintinger}},
  \bibinfo{author}{\bibfnamefont{S.}~\bibnamefont{Klein}},
  \bibinfo{author}{\bibfnamefont{F.}~\bibnamefont{Bustos}},
  \bibinfo{author}{\bibfnamefont{W.~L.} \bibnamefont{Barnes}},
  \bibnamefont{and} \bibinfo{author}{\bibfnamefont{T.~W.}
  \bibnamefont{Ebbesen}}, \bibinfo{journal}{Phys. Rev. B}
  \textbf{\bibinfo{volume}{71}}, \bibinfo{pages}{035424}
  (\bibinfo{year}{2005}).

\bibitem[{\citenamefont{Gupta et~al.}(2002)\citenamefont{Gupta, Dyer, and
  Weimer}}]{Gupta2002}
\bibinfo{author}{\bibfnamefont{R.}~\bibnamefont{Gupta}},
  \bibinfo{author}{\bibfnamefont{M.~J.} \bibnamefont{Dyer}}, \bibnamefont{and}
  \bibinfo{author}{\bibfnamefont{W.~A.} \bibnamefont{Weimer}},
  \bibinfo{journal}{J. Appl. Phys.} \textbf{\bibinfo{volume}{92}},
  \bibinfo{pages}{5264} (\bibinfo{year}{2002}).

\bibitem[{\citenamefont{Selwyn and Steinfeld}(1972)}]{Selwyn1972}
\bibinfo{author}{\bibfnamefont{J.~E.} \bibnamefont{Selwyn}} \bibnamefont{and}
  \bibinfo{author}{\bibfnamefont{J.~I.} \bibnamefont{Steinfeld}},
  \bibinfo{journal}{J. Phys. Chem.} \textbf{\bibinfo{volume}{76}},
  \bibinfo{pages}{762} (\bibinfo{year}{1972}).

\bibitem[{\citenamefont{Glass et~al.}(1980)\citenamefont{Glass, Liao, Bergman,
  and Olson}}]{Glass1980}
\bibinfo{author}{\bibfnamefont{A.~M.} \bibnamefont{Glass}},
  \bibinfo{author}{\bibfnamefont{P.~F.} \bibnamefont{Liao}},
  \bibinfo{author}{\bibfnamefont{J.~G.} \bibnamefont{Bergman}},
  \bibnamefont{and} \bibinfo{author}{\bibfnamefont{D.~H.} \bibnamefont{Olson}},
  \bibinfo{journal}{Opt. Lett.} \textbf{\bibinfo{volume}{5}},
  \bibinfo{pages}{368} (\bibinfo{year}{1980}).

\bibitem[{\citenamefont{Wurtz et~al.}(2007)\citenamefont{Wurtz, Evans, Hendren,
  Atkinson, Dickson, Pollard, Zayats, Harrison, and Bower}}]{Wurtz2007}
\bibinfo{author}{\bibfnamefont{G.~A.} \bibnamefont{Wurtz}},
  \bibinfo{author}{\bibfnamefont{P.~R.} \bibnamefont{Evans}},
  \bibinfo{author}{\bibfnamefont{W.}~\bibnamefont{Hendren}},
  \bibinfo{author}{\bibfnamefont{R.}~\bibnamefont{Atkinson}},
  \bibinfo{author}{\bibfnamefont{W.}~\bibnamefont{Dickson}},
  \bibinfo{author}{\bibfnamefont{R.~J.} \bibnamefont{Pollard}},
  \bibinfo{author}{\bibfnamefont{A.~V.} \bibnamefont{Zayats}},
  \bibinfo{author}{\bibfnamefont{W.}~\bibnamefont{Harrison}}, \bibnamefont{and}
  \bibinfo{author}{\bibfnamefont{C.}~\bibnamefont{Bower}},
  \bibinfo{journal}{Nano Lett.} \textbf{\bibinfo{volume}{7}},
  \bibinfo{pages}{1297} (\bibinfo{year}{2007}).

\bibitem[{\citenamefont{Markel et~al.}(1999)\citenamefont{Markel, Shalaev,
  Zhang, Huynh, Tay, Haslett, and Moskovits}}]{Markel1999}
\bibinfo{author}{\bibfnamefont{V.~A.} \bibnamefont{Markel}},
  \bibinfo{author}{\bibfnamefont{V.~M.} \bibnamefont{Shalaev}},
  \bibinfo{author}{\bibfnamefont{P.}~\bibnamefont{Zhang}},
  \bibinfo{author}{\bibfnamefont{W.}~\bibnamefont{Huynh}},
  \bibinfo{author}{\bibfnamefont{L.}~\bibnamefont{Tay}},
  \bibinfo{author}{\bibfnamefont{T.~L.} \bibnamefont{Haslett}},
  \bibnamefont{and}
  \bibinfo{author}{\bibfnamefont{M.}~\bibnamefont{Moskovits}},
  \bibinfo{journal}{Phys. Rev. B} \textbf{\bibinfo{volume}{59}},
  \bibinfo{pages}{10903} (\bibinfo{year}{1999}).

\bibitem[{\citenamefont{Norris}(1995)}]{Norris}
\bibinfo{author}{\bibfnamefont{T.~B.} \bibnamefont{Norris}},
  \bibnamefont{in} \emph{\bibinfo{title}{Confined Electrons and Photons}},
  \bibnamefont{edited by} \bibinfo{editor}{E.~Burstein and C.~Weisbuch}
  (\bibinfo{publisher}{Plenum, New York}, \bibinfo{year}{1995}), pp.
  \bibinfo{pages}{503--521}.

\bibitem[{\citenamefont{Noguez}(2007)}]{Noguez2007}
\bibinfo{author}{\bibfnamefont{C.}~\bibnamefont{Noguez}}, \bibinfo{journal}{J.
  Phys. Chem. C} \textbf{\bibinfo{volume}{111}}, \bibinfo{pages}{3806}
  (\bibinfo{year}{2007}).

\bibitem[{\citenamefont{Campion and Kambhampati}(1998)}]{campion98}
\bibinfo{author}{\bibfnamefont{A.}~\bibnamefont{Campion}} \bibnamefont{and}
  \bibinfo{author}{\bibfnamefont{P.}~\bibnamefont{Kambhampati}},
  \bibinfo{journal}{Chem. Soc. Rev.} \textbf{\bibinfo{volume}{27}},
  \bibinfo{pages}{241} (\bibinfo{year}{1998}).

\bibitem[{\citenamefont{Moskovits}(2005)}]{Moskovits2005}
\bibinfo{author}{\bibfnamefont{M.}~\bibnamefont{Moskovits}},
  \bibinfo{journal}{J. Raman Spectrosc.} \textbf{\bibinfo{volume}{36}},
  \bibinfo{pages}{485} (\bibinfo{year}{2005}).

\bibitem[{\citenamefont{Felidj et~al.}(2003)\citenamefont{Felidj, Aubard, Levi,
  Krenn, Hohenau, Schider, Leitner, and Aussenegg}}]{Felidj2003}
\bibinfo{author}{\bibfnamefont{N.}~\bibnamefont{Felidj}},
  \bibinfo{author}{\bibfnamefont{J.}~\bibnamefont{Aubard}},
  \bibinfo{author}{\bibfnamefont{G.}~\bibnamefont{Levi}},
  \bibinfo{author}{\bibfnamefont{J.~R.} \bibnamefont{Krenn}},
  \bibinfo{author}{\bibfnamefont{A.}~\bibnamefont{Hohenau}},
  \bibinfo{author}{\bibfnamefont{G.}~\bibnamefont{Schider}},
  \bibinfo{author}{\bibfnamefont{A.}~\bibnamefont{Leitner}}, \bibnamefont{and}
  \bibinfo{author}{\bibfnamefont{F.~R.} \bibnamefont{Aussenegg}},
  \bibinfo{journal}{Appl. Phys. Lett.} \textbf{\bibinfo{volume}{82}},
  \bibinfo{pages}{3095} (\bibinfo{year}{2003}).

\bibitem[{\citenamefont{McFarland et~al.}(2005)\citenamefont{McFarland, Young,
  Dieringer, and Van~Duyne}}]{McFarland2005}
\bibinfo{author}{\bibfnamefont{A.~D.} \bibnamefont{McFarland}},
  \bibinfo{author}{\bibfnamefont{M.~A.} \bibnamefont{Young}},
  \bibinfo{author}{\bibfnamefont{J.~A.} \bibnamefont{Dieringer}},
  \bibnamefont{and} \bibinfo{author}{\bibfnamefont{R.~P.}
  \bibnamefont{Van~Duyne}}, \bibinfo{journal}{J. Phys. Chem. B}
  \textbf{\bibinfo{volume}{109}}, \bibinfo{pages}{11279}
  (\bibinfo{year}{2005}).

\bibitem[{\citenamefont{Persson et~al.}(2006)\citenamefont{Persson, Zhao, and
  Zhang}}]{Persson2006}
\bibinfo{author}{\bibfnamefont{B.~N.~J.} \bibnamefont{Persson}},
  \bibinfo{author}{\bibfnamefont{K.}~\bibnamefont{Zhao}}, \bibnamefont{and}
  \bibinfo{author}{\bibfnamefont{Z.}~\bibnamefont{Zhang}},
  \bibinfo{journal}{Phys. Rev. Lett.} \textbf{\bibinfo{volume}{96}},
  \bibinfo{pages}{207401} (\bibinfo{year}{2006}).

\bibitem[{\citenamefont{Pettinger}(1986)}]{Pettinger1986}
\bibinfo{author}{\bibfnamefont{B.}~\bibnamefont{Pettinger}},
  \bibinfo{journal}{J. Chem. Phys.} \textbf{\bibinfo{volume}{85}},
  \bibinfo{pages}{7442} (\bibinfo{year}{1986}).

\end{thebibliography}
\end{document}